\documentclass[prd,aps,twocolumn,a4paper,floatfix,showpacs,nofootinbib]%
{revtex4-1}

\usepackage{graphicx,psfrag}
\usepackage[printwatermark]{xwatermark}
\usepackage{xcolor}
\usepackage{mathrsfs}
\usepackage{amsmath,amsfonts,amssymb,pifont,bm}
\usepackage{multirow,enumerate}
\usepackage{comment,hyperref}
\usepackage{color}
\usepackage{acronym}
\usepackage{xspace}
\usepackage[normalem]{ulem}
\usepackage{xfrac}
\usepackage[group-digits=false]{siunitx}
\usepackage{booktabs,makecell,multirow}
\usepackage{txfonts}
\usepackage{soul}
\usepackage{lineno}
\usepackage{enumitem}

\newacro{BH}{black hole}
\newacro{NS}{neutron star}
\newacro{PN}{post-Newtonian}
\newacro{BBH}{binary-black-hole}
\newacro{BNS}{binary neutron star}
\newacro{NSBH}{neutron-star black-hole}
\newacro{EOB}{effective one-body}
\newacro{NR}{numerical relativity}
\newacro{GW}{gravitational-wave}
\newacro{PSD}{power spectral density}
\newacro{aLIGO}{advanced Laser Interferometer Gravitational-wave Observatory}
\newacro{AZDHP}{aLIGO zero detuned high power density}
\newacro{GR}{general relativity}
\newacro{PE}{parameter estimation}
\newacro{LAL}{LIGO algorithm library}
\newacro{GPR}{Gaussian process regression}
\newacro{ROM}{reduced-order model}
\newacro{IMR}{inspiral-merger-ringdown}
\newacro{CE}{Cosmic Explorer}
\newacro{ET}{Einstein Telescope}
\newacro{WD}{white dwarf}
\newacro{SNR}{signal to noise ratio}

\newcommand{\be}{\begin{equation}}
\newcommand{\ee}{\end{equation}}
\newcommand{\bea}{\begin{eqnarray}}
\newcommand{\eea}{\end{eqnarray}}
\newcommand{\bel}{\begin{align}}
\newcommand{\eel}{\end{align}}

\newcommand\numberthis{\addtocounter{equation}{1}\tag{\theequation}}

\def\GMc2{{\rm G M_{\odot} c^{-2}}}

\usepackage{color}
\definecolor{cyan}{rgb}{0,0.9,0.9}
\definecolor{orange}{rgb}{0.9,0.5,0}
\definecolor{magenta}{rgb}{1,0,1}
\definecolor{purple}{rgb}{0.8,0.4,0.8}
\definecolor{gray}{rgb}{0.5,0.5,0.5}
\definecolor{mygreen}{rgb}{0.1,0.8,0.1}
\definecolor{darkblue}{rgb}{0.0,0.0,0.6}

\newcommand{\pyrex}{\texttt{pyrex}\xspace}

\newlist{subenum}{enumerate}{1}
\setlist[subenum,1]{label=(\alph*)}

\newcommand{\AEIHannover}{Max Planck Institute for Gravitational Physics
(Albert Einstein Institute), Callinstra{\ss}e~38, 30167 Hannover, Germany}
\newcommand{\UniHannover}{Leibniz Universit\"at Hannover, 30167 Hannover, 
Germany}
\newcommand{\GRASP}{Institute for Gravitational and Subatomic Physics (GRASP)
Department of Physics, Utrecht University, Princetonplein 1, 3584 CC Utrecht, The Netherlands}

\begin{document}

\title{Adding eccentricity to quasicircular binary-black-hole waveform models}

\author{Yoshinta \surname{Setyawati}}
\affiliation{\AEIHannover}
\affiliation{\UniHannover}
\affiliation{\GRASP}

\author{Frank \surname{Ohme}}

\affiliation{\AEIHannover}
\affiliation{\UniHannover}

\date{\today}

\begin{abstract}
The detection of gravitational-wave signals from 
coalescing eccentric binary black holes would yield unprecedented information 
about the formation and evolution of compact binaries in specific scenarios, 
such as dynamical formation in dense stellar clusters and
three-body interactions.
The gravitational-wave searches by the ground-based interferometers, LIGO and 
Virgo, rely on analytical waveform models for binaries on quasicircular orbits.
Eccentric merger waveform models are less developed, and only few 
numerical simulations of eccentric mergers are publicly available, but
several eccentric inspiral models have been developed from the Post-Newtonian 
expansion. 
Here we present a novel method to convert the dominant quadrupolar mode of any circular analytical binary-black-hole model 
into an eccentric model.
First, using numerical simulations, we examine the additional amplitude and 
frequency modulations of eccentric signals that are not present in their 
circular counterparts. 
Subsequently, we identify suitable analytical descriptions of those 
modulations and interpolate key parameters from twelve numerical simulations 
designated as our training dataset. This allows us
to reconstruct the modulated amplitude and phase of any waveform up to mass ratio 3 and eccentricity 0.2.
We find that the minimum overlap of the new model with numerical simulations is 
around 0.98 over all of our test dataset that are scaled to a 50M$_\odot$ black-hole binary starting at 35 Hz with aLIGO A+ design sensitivity.
A Python package \pyrex easily carries out the computation of this method.
 
\end{abstract}

\maketitle


\section{Introduction}
\label{sec:intro}

Coalescing stellar-mass black-hole binaries are one of the primary sources of \ac{GW} signals detected by the ground-based interferometers, the \ac{aLIGO} \cite{advancedLIGO}, Virgo \cite{advanceVirgo}, and KAGRA \cite{kagra}.
In the first three observing runs (O1--O3), detection pipelines assumed \ac{BBH} 
mergers to have negligible eccentricity when entering the orbital 
frequencies to which \ac{aLIGO}, Virgo, and KAGRA are sensitive 
\cite{PhysRevX.9.031040, LIGOeccen, GWTC2}. 
\acp{BBH} formed in an isolated environment through a massive stellar evolution are expected to circularize and therefore have undetectable eccentricity by the time they enter the LIGO band \cite{PhysRev.136.B1224}.
However, \acp{BBH} with a detectable eccentricity can form in a dense stellar cluster through dynamical capture \cite{Samsing_2014, PhysRevD.98.083028}.

A possible scenario is that the binary gains eccentricity due to gravitational 
torques exchanged with a circumbinary disk \cite{refId0}.
Eccentric \acp{BBH} can also form from three-body interactions 
\cite{PhysRevD.98.083028}, where the \ac{BBH} behaves as the inner binary.  
In this system, the Kozai-Lidov \cite{kozai, lidov} mechanism triggers the oscillation that boosts the eccentricity.

Interactions of \acp{BBH} in a typical globular cluster suggest a significant eccentric \ac{BBH} merger rate.  As many as $\sim$$5\%$ of binaries may enter the LIGO detector band ($f \geq$ 10 Hz) with eccentricities $e> 0.1$ \cite{gloclus, PhysRevD.98.123005, PhysRevD.97.103014}. 
A confident measurement of significant eccentricity in a \ac{BBH} system would be strong evidence for the dynamical formation scenarios in dense 
stellar clusters and would boost our understanding of the dynamical evolution of compact objects.

The impact of eccentricity is more substantial during the early 
inspiral and therefore plays a vital role in the space-based detector era 
\cite{lisa}.
In the LIGO band, the detection of \acp{GW} from an eccentric orbit would 
suggest that the binary was formed with a small initial separation and 
did not have time to circularize, or the binary evolved through an unknown 
dynamical process.
Incorporating eccentric \ac{BBH} simulations may also lead to an 
increase in the LIGO/Virgo/KAGRA detection rate \cite{PhysRevD.98.123005}.
Besides, the detection of eccentric \ac{BBH} mergers could capture effects from the extreme-gravity regime and therefore can be used for testing the general theory of relativity \cite{PhysRevD.100.124032, YunesLiv}.
   
We highlight the significance of detecting \acp{GW} from eccentric \acp{BBH}.
Constructing template models for eccentric waveforms is challenging, and we aim to make progress towards this goal especially for the late inspiral and merger regimes that are most accessible with today's observations.
One of the main difficulties in developing an eccentric waveform model is that 
only a few \ac{NR} simulations with higher eccentricity are available. 
Thus, many studies focus on developing eccentric models from the \ac{PN} 
expansion. 
The development of full \ac{IMR} eccentric waveform models is currently an 
actively researched area \cite{PhysRevD.97.024031, PhysRevD.96.044028, 
PhysRevD.98.044015}.

Huerta \textit{et} al. \cite{PhysRevD.97.024031} construct a time-domain eccentric nonspinning waveform model ($e_0<0.2$) up to mass ratio 5.5, where $e_0$ is the eccentricity 10 cycles before the merger. 
Their model is called {\texttt{ENIGMA}}, a hybrid waveform that has been 
calibrated using a set of numerical simulations and trained using \ac{GPR}.
Reference~\cite{PhysRevD.96.044028} presents a low-eccentricity model ($e_0<0.2$) 
called {\texttt{SEOBNRE}} using the expansion of the \ac{EOB} waveform family.
A more up-to-date \ac{EOB} formalism is demonstrated in Refs.~\cite{PhysRevD.101.101501,Nagar:2021gss}.
Hinder \textit{et} al.~\cite{PhysRevD.98.044015} present a time-domain, nonspinning 
eccentric waveform model up to mass ratio $q=m_1/m_2=3$ from 23 \ac{NR} 
simulations that are publicly available in the SXS catalog.
The referenced eccentricity is $e_{\textrm{ref}} \leq 0.08$ starting at seven cycles before the merger. 
Like Ref.~\cite{PhysRevD.97.024031}, the early inspiral of this model is 
hybridized with a \ac{PN} expansion to produce a full \ac{IMR} model in a 
\textit{Mathematica} package \cite{mathhinder}.
In addition, Ref.~\cite{PhysRevD.103.064022} recently developed an eccentric model {\texttt{NRSur2dq1Ecc}} for nonspinning waveforms and eccentricities up to 0.2 from 47 \ac{NR} simulations.
Although the model was trained for $q=1$, it can be extended to mass ratio $q \approx 3$.
Apart from the studies above, nonspinning, low-eccentricity frequency-domain 
models from the \ac{PN} expansion are publicly available in the \ac{LAL} 
\cite{PhysRevD.93.064031, PhysRevD.90.084016, PhysRevD.93.124061}.

The excitement to search for an eccentric \ac{BBH} motivated the following analysis.
References~\cite{10.1093/mnras/stz2996, 10.1093/mnrasl/slaa084, Romero_Shaw_2020} recently developed an analysis to find the signature of an eccentric \ac{BBH} in the O1, O2 and several events in the O3 data using the {\texttt{SEOBNRE}} model.
Additionally, Ref.~\cite{gayathri2020gw190521} analyzed the heaviest \ac{BBH} system during O1--O3, GW190521 \cite{PhysRevLett.125.101102} with 325 \ac{NR} simulations.
They found that this event is consistent with highly precessing, eccentric model with $e \approx 0.7$.

We present a promising method to add eccentricity to quasicircular systems independent of the \ac{PN} expansion.
We apply this method to nonspinning, time-domain waveforms, although in principle it can be used in more general settings. 
Our technique focuses on a fast reconstruction of the near-merger eccentric 
\ac{BBH} waveform and can be applied to any analytical circular nonspinning 
model.
We build our model from 12 \ac{NR} simulations and test against further 8 
\ac{NR} simulations from the open SXS catalog \cite{SXS}.
Our method is very simple and can be applied to \textit{any} circular time-domain model 
obtained from, e.g., the phenomenological \cite{PhysRevD.93.044007, 
Hannam:2013oca, PhysRevD.82.064016} or \ac{EOB} \cite{PhysRevD.59.084006, 
PhysRevD.81.084041} families.

We model the deviation from circularity visible in the amplitude and phase of eccentric GW signals. 
This deviation is modeled across the parameter space and can be simply added to any quasicircular model, which elevates that model to include eccentric effects.
This approach is inspired by the "twisting" technique that is applied for 
reconstructing precessing spins from an aligned-spin 
model to build, e.g., the {\texttt{IMRPhenomP}} family
\cite{Hannam:2013oca,2020PhRvD.101b4056K,Khan:2018fmp,Pratten:2020ceb,
Estelles:2020osj}.
The dynamic calibration of the waveform model is motivated by our previous study 
\cite{PhysRevD.99.024010} and the regression techniques tested in detail in Ref. ~\cite{Setyawati_2020}.

We calibrate our model for mass ratios $q \leq 3$ and eccentricity $e \leq 
0.2$, and provide it as a Python package called \pyrex 
\cite{pyrexzen}.
Our model has been constructed for a fiducial 50 $M_\odot$ \ac{BBH} and 
can then be rescaled for other total masses $M$.
We find that the overlap of all our test data against \ac{NR} is above 98\%.
Moreover, we expand the construction to earlier regimes than the calibrated time span. 
Although we do not calibrate for higher mass ratios, the early inspiral, or higher orbital eccentricity, we allow the building of waveforms beyond the parameter boundaries used for calibration.

The organization of this manuscript is as follows:
In Sec.~\ref{sec:method}, we present the methodology to construct 
this model.
Section \ref{sec:result} discusses the primary outcome and the faithfulness of 
our model.
Finally, Sec.~\ref{sec:conclusion} summarizes and concludes the prospect of our studies.
Throughout this article, we use geometric units in which $G=c=1$.
 

\section{Method}
\label{sec:method}

Using \ac{NR} simulations, we investigate the frequency and amplitude 
modulations in eccentric \ac{BBH} 
signals and implement them in analytical waveforms to 
develop our model.
As described by Peters \cite{PhysRev.136.B1224}, the orbital eccentricity 
in binary systems decreases over time due to energy loss through \ac{GW} 
radiation.
Pfeiffer \textit{et} al.~\cite{Pfeiffer_2007} investigated this in numerical 
simulations of the SXS catalog.
The authors point out that one of the main differences in the evolution of low-eccentricity initial data compared to quasicircular binaries is an overall time and 
phase shift, where the quasicircular data represent the binary at a point close to merger.
Following these studies, Hinder et al.~\cite{PhysRevD.98.044015} showed that the 
\ac{GW} emissions from low-eccentric binaries and circular binaries are
indistinguishable near the merger stage.
Specifically, Hinder \textit{et} al.\ suggest that one only loses 4\% of the signal when 
substituting the \ac{GW} emission from low-eccentricity binaries with circular 
orbits 30$M$ before the peak of the amplitude ($t=0$).
They use this fact to build an eccentric \ac{IMR} model by replacing the late inspiral eccentric model with a circular waveform.
Combining the finding above, we model the decaying eccentricity as amplitude and 
phase modulation up to $t = -29M$.
We then substitute the \ac{GW} strain at $t>-29M$ with the circular model for the 
same binary masses.

\subsection{Data preparation}
\label{sec:methodprepare}

\begin{figure}
\centering
\includegraphics[width=\hsize]{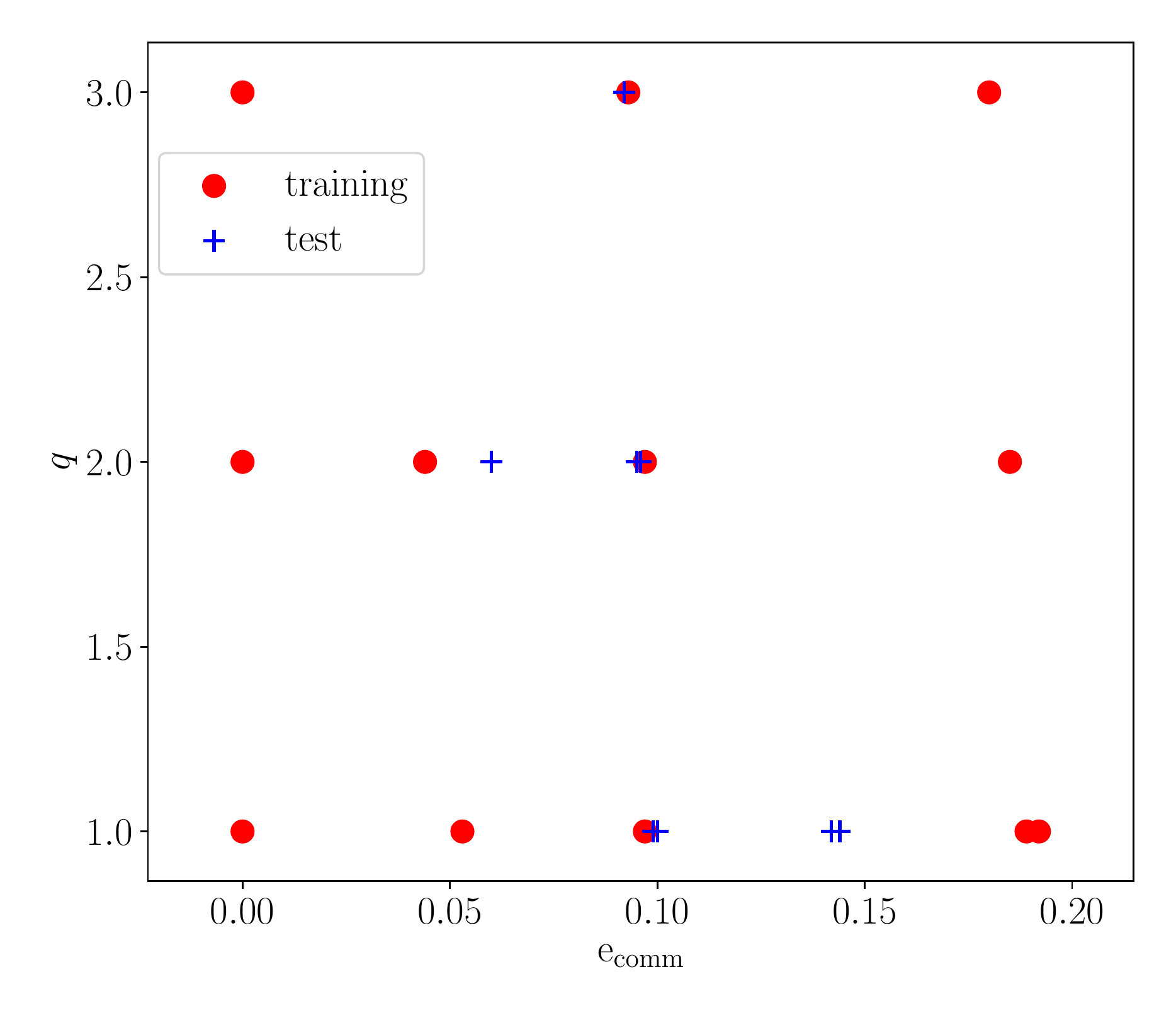}
\caption{The training and test data, shown by the red circles and the blue plus 
signs, are located in the parameter space of mass ratio and 
eccentricity. 
We use 20 \ac{NR} simulations from the SXS catalog and divide them into 12 \ac{NR} training datasets and 8 test datasets.}
\label{fig:nr}
\end{figure}

We use 20 nonspinning \ac{NR} simulations from the SXS catalog up to mass ratio 3 and eccentricity 0.2 to build our model (see Table \ref{tab:sxsdata}).
We follow the definition of eccentricity $e_\textrm{comm}$ in Ref.~\cite {PhysRevD.98.044015} as the eccentricity measured at the referenced frequency, $x=(M\omega)^{2/3}=0.075$. 
These simulations are divided into a training data set of 12 simulations and 
the test datasets of 8 simulations, as shown in Fig.~\ref{fig:nr}. Binaries
of the test dataset fall within the training data's parameter boundaries.
Hence, we do not perform extrapolation with the test data.

We combine the ``+'' and ``$\times$'' polarization using the spin-weighted 
spherical harmonics with the following expression \cite{formatNR}: 
\be
\label{eq:NRform}
h_{+}-ih_\times=\frac{M}{r}\sum_{\ell=2}^{\infty} \sum_{m=-\ell}^{m=\ell}h_{\ell 
m}(t) \; ^{-2}Y_{\ell m}(\iota, \phi),
\ee
where $M$ and $r$ are the total mass of the system and the distance from the 
observer, respectively; $^{-2}Y_{\ell m}$ are the spin-weighted spherical 
harmonics that depend on the inclination angle $\iota$ and the phase angle 
$\phi$; and $h_{\ell m}(t)$ can be extracted from the \ac{NR} 
data in the corresponding catalog.  
We construct our model for $h_{2 \pm 2}$, the leading contribution of spherical 
harmonic modes with $\ell=2$, $m=\pm 2$. 
Reference~\cite{PhysRevD.98.044015} suggests that other, subdominant modes are less 
significant for nearly equal-mass systems with low eccentricity.
Here we consider only moderately small eccentricities; therefore we only model 
the dominant mode.
For future studies, subdominant harmonics will be important to model high-eccentricity signals accurately.

\begin{table}[t]
\centering
\caption{\ac{NR} simulations from the SXS catalog used in this study with mass 
ratio $q=m_1/m_2$, eccentricity at the reference frequency $e_{\textrm{comm}}$, 
and the number of orbits before the maximum amplitude of $\|{h_\textrm{22}}\|$. 
$e_\textrm{comm}$ is the eccentricity at the reference frequency $(M\omega)^{2/3}\,=\,0.075$ as described in Ref.~\cite{PhysRevD.98.044015}.
The quasicircular waveforms ($e_{\textrm{comm}}$\,=\,0.000) have eccentricities lower than 10$^{-5}$ at the reference frequency.}
\label{tab:sxsdata}
\begin{ruledtabular}
\begin{small}
   \begin{tabular}{llllll}
    Case & Simulations & Training/test & $q$ & $e_{\textrm{comm}}$ & $N_{\textrm{orbs}}$\\
   \hline
  1& SXS:BBH:0180 & Training & 1 & 0.000 & 26.7 \\
  2 & SXS:BBH:1355 & Training & 1 & 0.053 & 11.9 \\
  3 & SXS:BBH:1357 & Training & 1 & 0.097 &12.8 \\
  4 & SXS:BBH:1358 & Test & 1 & 0.099 &12.1 \\
  5 & SXS:BBH:1359 & Test & 1 &  0.100 & 11.7 \\
  6 & SXS:BBH:1360 & Test & 1 & 0.142 &11.1\\
  7 & SXS:BBH:1361 & Test & 1 & 0.144 &10.9 \\
  8 & SXS:BBH:1362 & Training & 1 & 0.189 &10.2 \\
  9 & SXS:BBH:1363 & Training & 1 & 0.192 &10.1 \\
  10 & SXS:BBH:0184 & Training & 2 & 0.000 & 13.7\\
  11 & SXS:BBH:1364 & Training & 2 & 0.044 & 14.2\\
  12 & SXS:BBH:1365 & Test & 2 & 0.060 & 14.1 \\
  13 & SXS:BBH:1366 & Test & 2 & 0.095 & 13.6 \\
  14 & SXS:BBH:1367 & Test & 2 & 0.096 & 13.6 \\
  15 & SXS:BBH:1368 & Training & 2 & 0.097 & 13.6 \\
  16 & SXS:BBH:1369 & Training & 2 & 0.185 & 13.6 \\
  17 & SXS:BBH:0183 & Training & 3 & 0.000 & 13.5 \\
  18 & SXS:BBH:1372 & Test & 3 & 0.092 & 15.6 \\
  19 & SXS:BBH:1373 & Training & 3 & 0.093 &15.3 \\
  20 & SXS:BBH:1374 & Training & 3 & 0.180 & 13.5 \\
\end{tabular}
\end{small}
\end{ruledtabular}
\end{table}

\begin{figure}
\includegraphics[width=\linewidth]{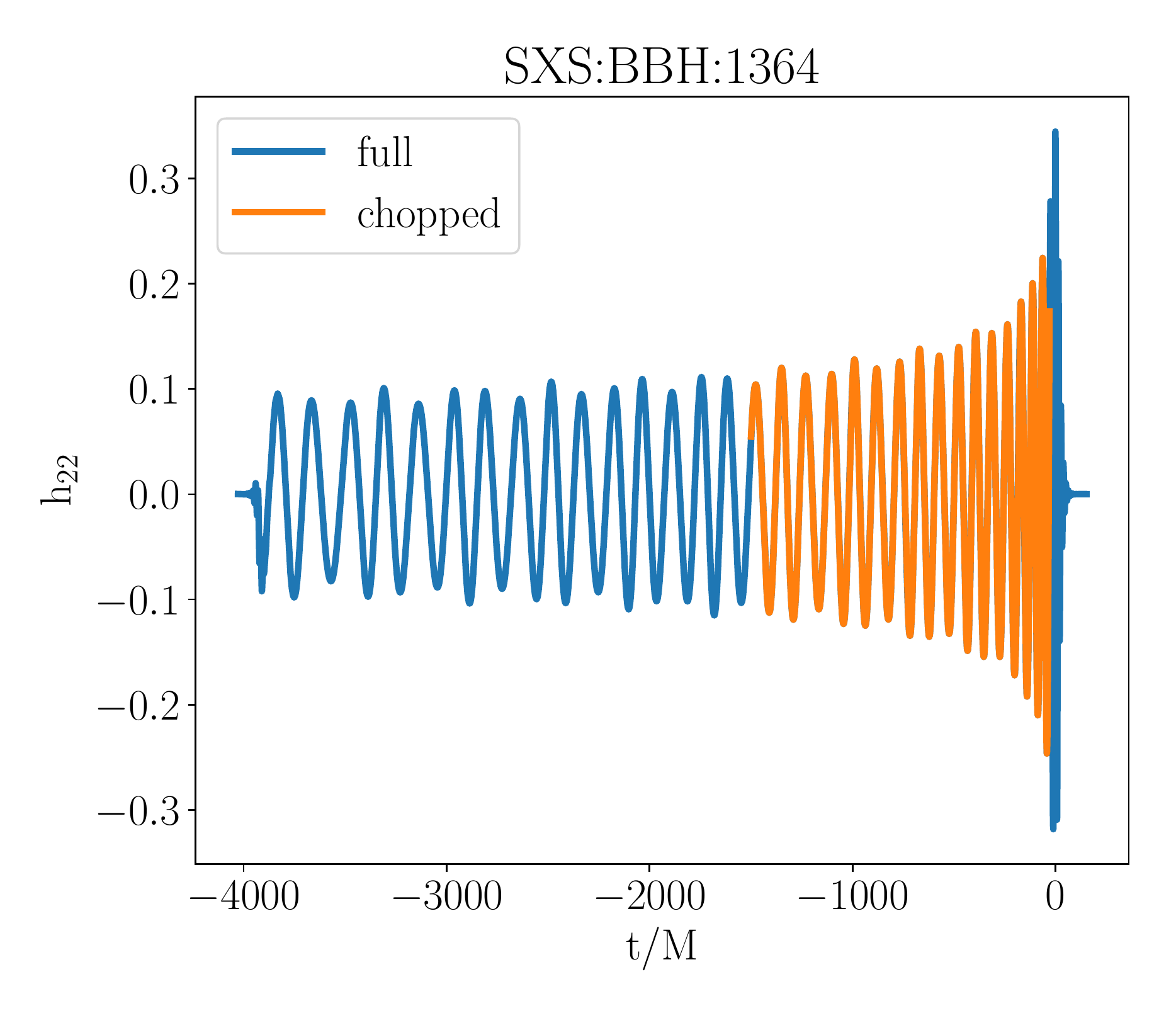}
\caption{The full and the chopped waveform of the SXS:BBH:1364 simulation ($q=2, e_{\textrm{comm}}=0.044$). The blue line shows the full \ac{NR} $h_{\textrm{22}}$ mode, and the orange line presents the time range used in this study. 
We remove the first $250M$ due to the junk radiation and modulate the residual oscillation at $-1500M \leq t \leq -29M$.}
\label{fig:waves}
\end{figure}

We prepare the data as follows.
First, we align all the waveforms in the time domain such that the peak amplitude is at $t=0$.
Subsequently, we remove the first 250$M$ from the start of the waveforms due to the junk radiation,
and the last 29$M$ before $t=0$ due to circularization (see Fig.~\ref{fig:waves}).
Later, we use a circular waveform for $t>-29M$.
We then decompose $h_{2 \pm 2}$ into amplitude ($\mathcal{A}$), phase ($\Psi$), and the phase derivative, $\omega=\frac{d \Psi}{dt}$, where the referenced frequency follows Ref~\cite{PhysRevD.98.044015}.

We model amplitude $\mathcal{A}_\textrm{22}$ and frequency ($\omega_\textrm{22}$) as a simple quasicircular piece plus an oscillatory function. 
The final model then yields the phase ($\Psi_\textrm{22}$) by integrating the frequency.

\begin{figure}
	\begin{minipage}{0.49\linewidth}
		\includegraphics[width=\linewidth]{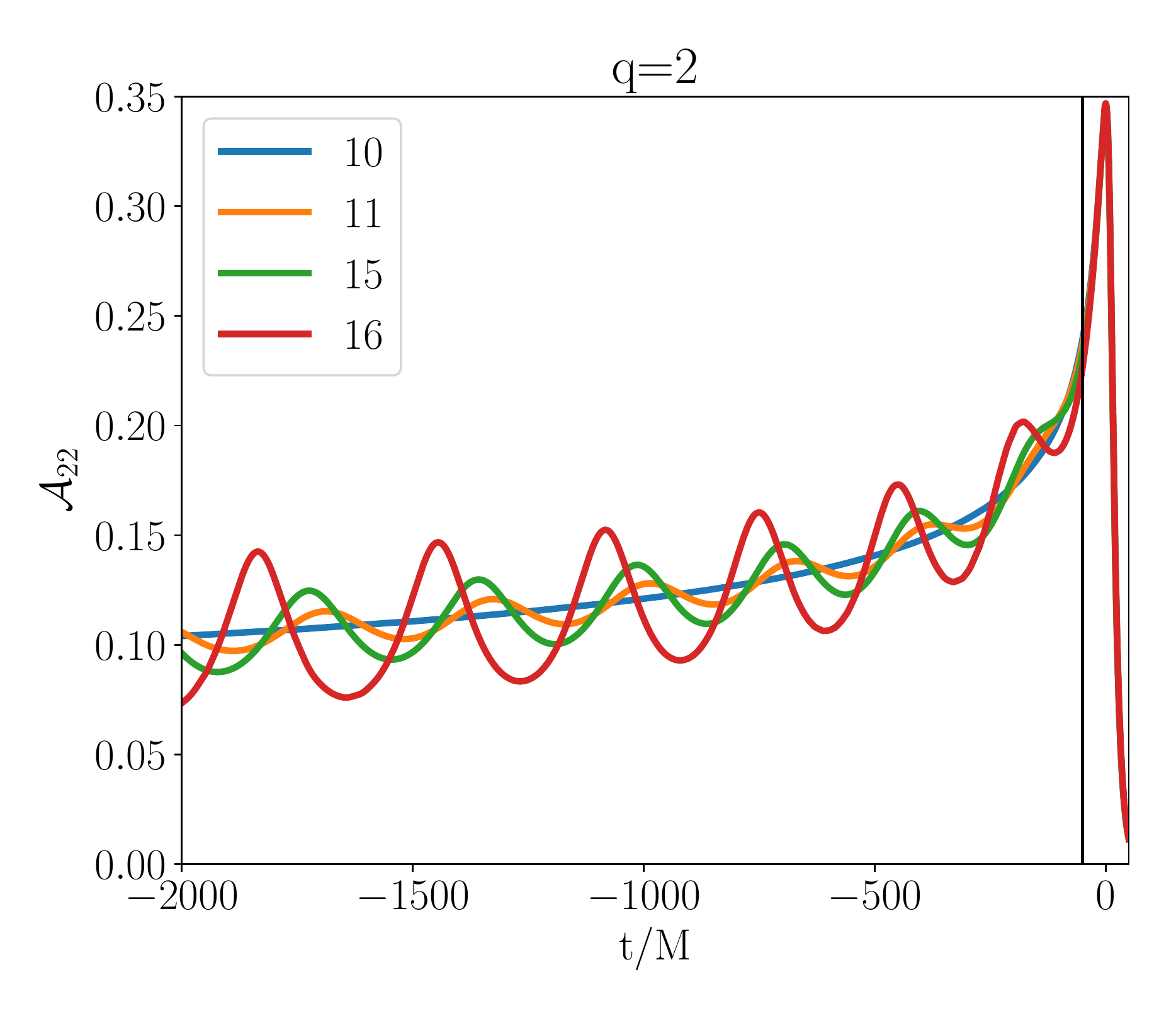}
	\end{minipage} \hfill
	\begin{minipage}{0.49\linewidth}
		\includegraphics[width=\linewidth]{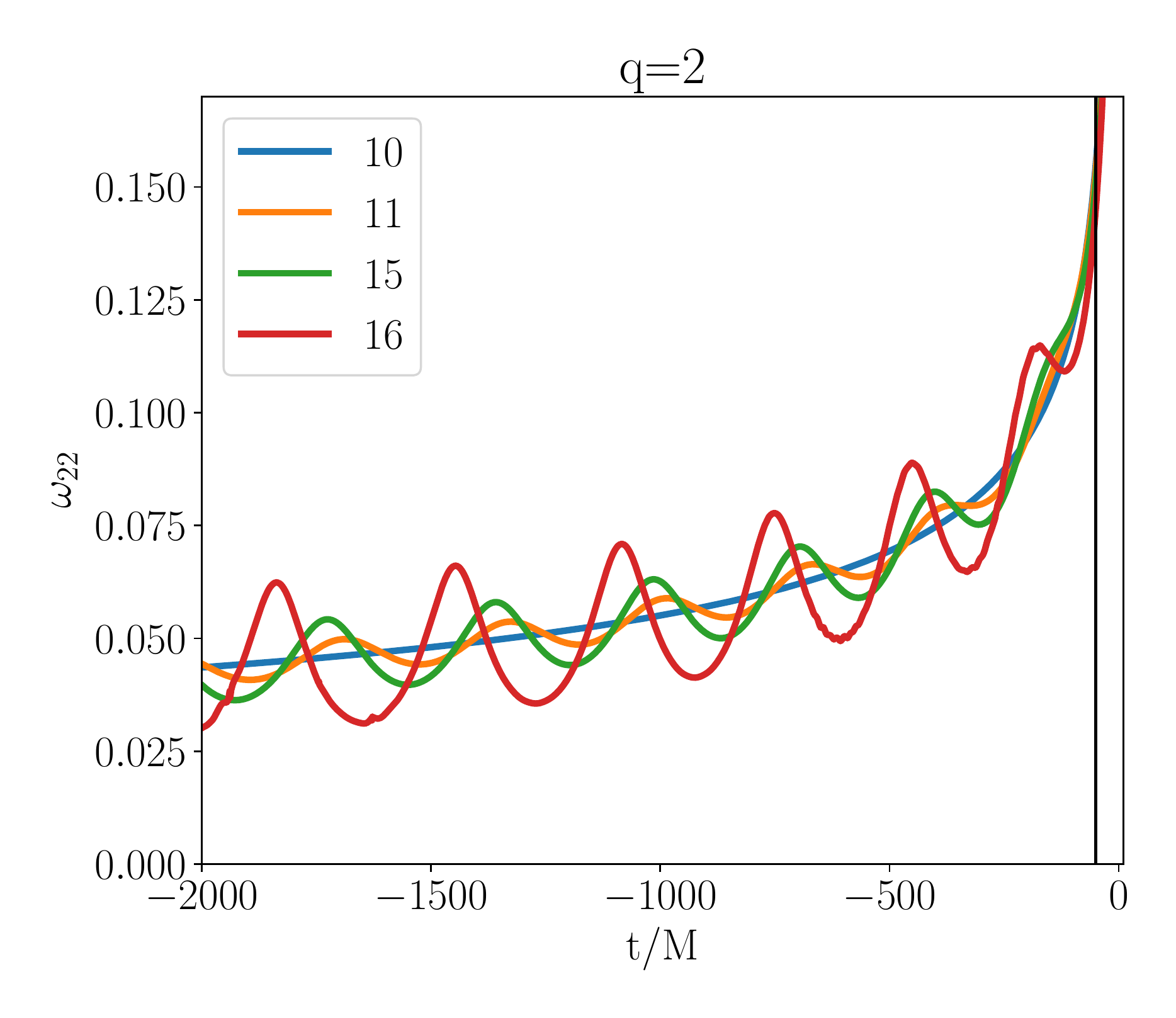}
	\end{minipage}
	\vspace{4ex}
	\begin{minipage}{0.49\linewidth}
		\includegraphics[width=\linewidth]{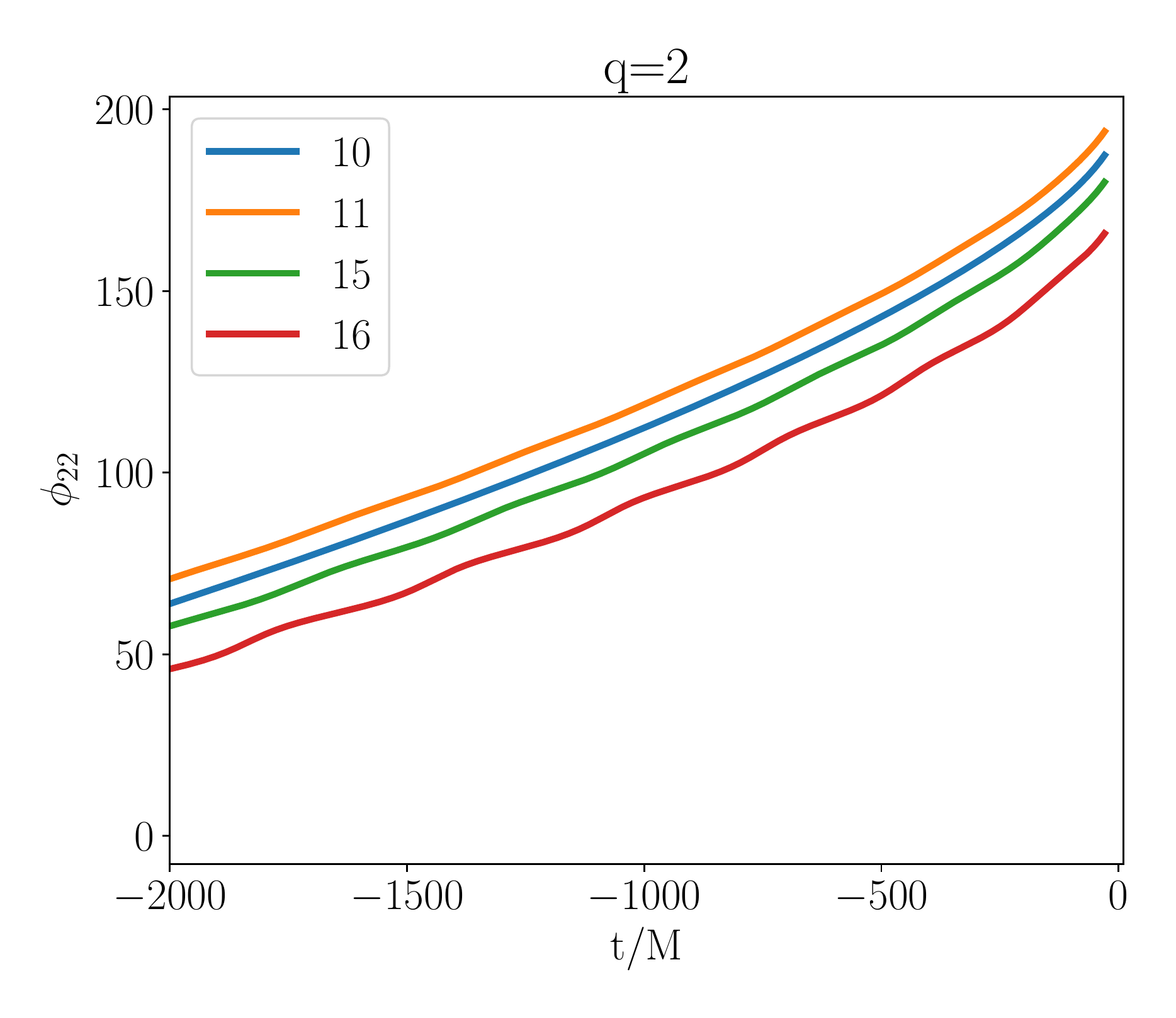}
	\end{minipage}
	\caption{The top-left panel shows the amplitude, the top-right 
panel shows the time derivative of the phase $\omega_{\textrm{22}}=d\Psi_{\textrm{22}}/dt$, and the bottom panel shows the phase of $h_\textrm{22}$.
We present the key parameters from the training dataset for $q=2$ ($\ell=2, m=2$). 
The numbers in the legend correspond to the 
case numbers of the simulations shown in Table~\ref{tab:sxsdata}. Although 
higher-eccentricity waveforms produce more oscillations than the lower-eccentricity waveforms, all data appear identical at $t>-30M$ due to 
circularization as shown in the top panels.
We employ the residual amplitude $\mathcal{A}_\textrm{22}$ and frequency $\omega_\textrm{22}$ to develop our model in the late inspiral regime.}
\label{fig:components}
\end{figure}

\subsection{Eccentricity estimator}
\label{sec:methodest}

In numerical simulations, eccentricity is often discussed as a consequence 
of imperfections in the initial data \cite{Ramos-Buades:2018azo}. 
It manifests itself as small oscillations on top of the gradual 
binary evolution, where the oscillation's amplitude is proportional to the 
eccentricity (see $\mathcal{A}_{\textrm{22}}$ and 
$\omega_{\textrm{22}}$ plots in Figs.~\ref{fig:waves} and \ref{fig:components}).
We use this residual oscillation as a key to estimating the eccentricity 
evolution. 

Mrou\'e \textit{et} al.~\cite{PhysRevD.82.124016} compare various methods to estimate 
eccentricity using $e_{\textrm{X}} (t)$.
The orbital eccentricity is proportional to the amplitude of a sinusoidal 
function, $e_{\textrm{X}} (t)$, expressed by %
\begin{align*}
 e_X (t) &=\frac{X_{\textrm{NR}}(t)-X_{\textrm{c}}(t)}{2 X_{\textrm{c}}(t)}, \\
\Leftrightarrow  e_X(X_c) &=\frac{X_{\textrm{NR}}(X_c)-X_{\textrm{c}}}{2 
X_{\textrm{c}}}, \numberthis \label{eq:estimator}
\end{align*}
where $X$ is either $\omega_{\textrm{22}}$ or $\mathcal{A}_{\textrm{22}}$, and 
$X_{\textrm{c}} (t)$ is the $X$ quantity in circular binaries instead of 
low-order polynomial fitting functions that are often used in the literature.
We reverse this relation to convert a circular model [with given $X_c(t)$] to an 
eccentric model using an analytical description of the oscillatory function 
$e_X(X_c)$.  
We apply the Savitzky-Golay filter \cite{savgol} to smooth the $e_X (t)$ curves from noises caused by numerical artifacts.
Savitzky-Golay is a digital filter applied to smooth the selected data points without altering the signal direction by fitting the adjacent data with a low-degree polynomial fit.

We stress that the definition of the orbital eccentricity is not unique. 
Thus, one could use different definitions of eccentricity. 
In principle, any definition can be accepted if consistently applied to the study in question.  
The \ac{NR} data we use are labeled with a value for the initial eccentricity that is based on \ac{PN} initial data \cite{PhysRevD.98.044015}. As we shall discuss below,
these labels are similar to what we estimate for the eccentricity using Eq.~\eqref{eq:estimator}, but not identical.
However, we refrain from redefinition of the initial eccentricity of the \ac{NR} data and instead identify each \ac{NR} simulation with the value of eccentricity at the reference frequency $(M\omega)^{2/3}=0.075$ determined by the original Ref.~\cite{PhysRevD.98.044015}. We do this because
(i) we want to avoid any confusion as to what \ac{NR} data we are using and what their properties are, and (ii)
by making the amplitude of $e_X$ a function of the eccentricity label imposed by Ref.~\cite{PhysRevD.98.044015}, we introduce an extra uncertainty that may 
be seen as representing the ambiguity in determining the initial 
eccentricity of the respected \ac{NR} simulations. Thus, we present a conservative 
estimate of the approach's accuracy.

\begin{figure}
	\begin{minipage}{0.49\linewidth}
		\includegraphics[width=\linewidth]{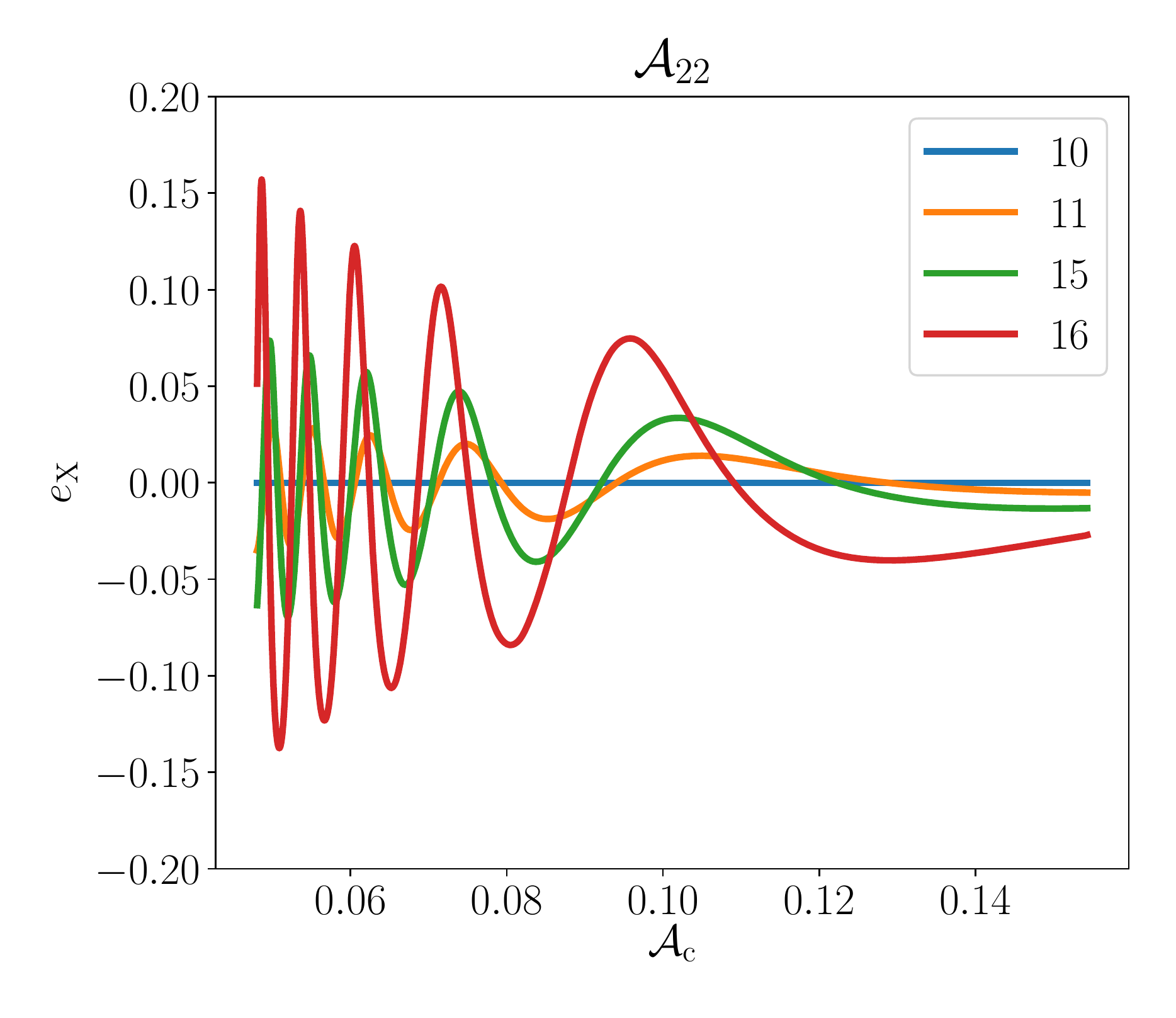}

	\end{minipage} \hfill
	\begin{minipage}{0.49\linewidth}
		\includegraphics[width=\linewidth]{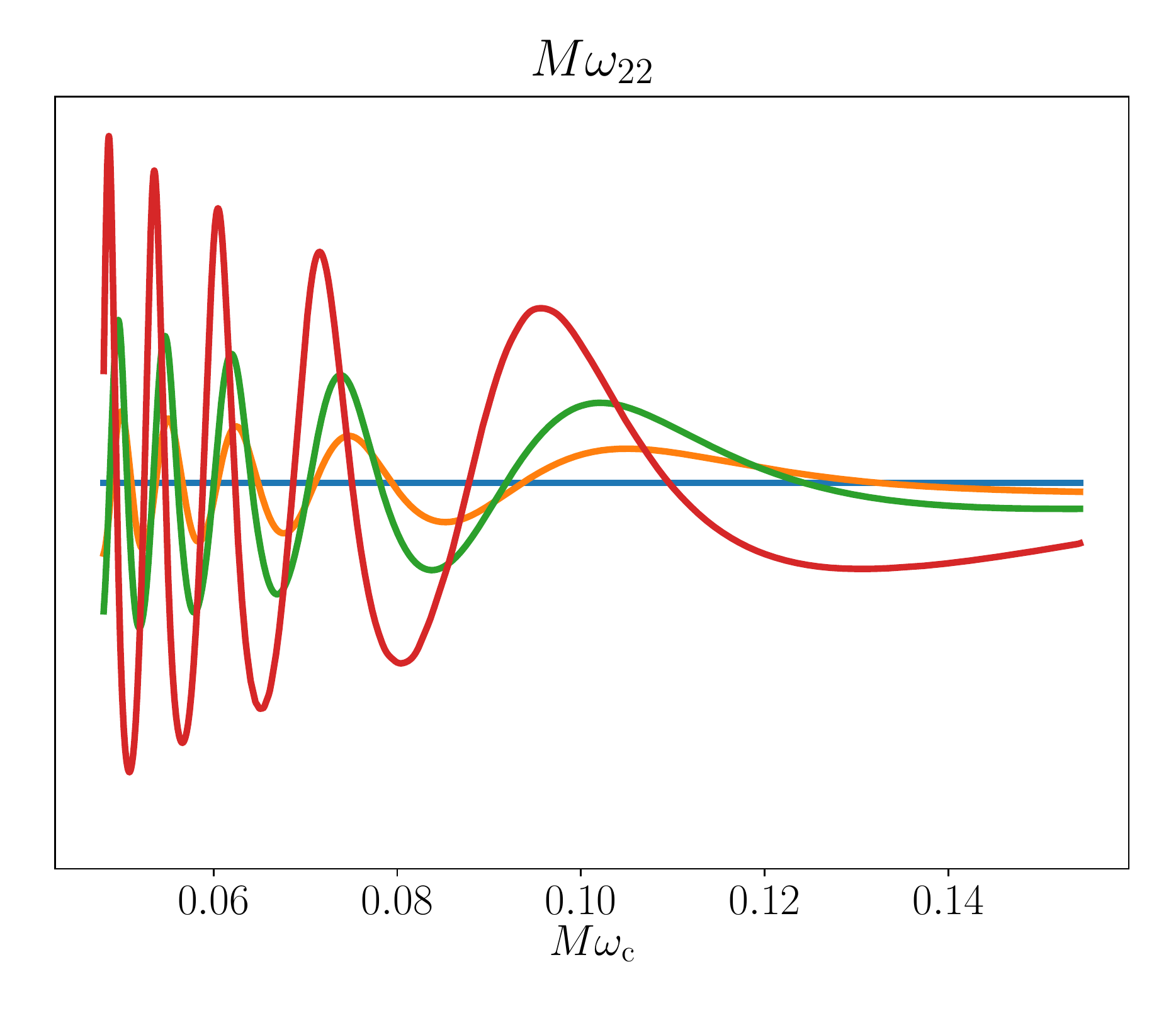}

	\end{minipage}
	\caption{The eccentricity estimator from $\mathcal{A}_{\textrm{22}}$ plotted 
against the circular amplitude $\mathcal{A}_{\textrm{c}}$ (left), and the
	eccentricity estimator from $\omega_{\textrm{22}}$ plotted against the 
circular omega $\omega_{\textrm{c}}$ (right) with the same mass ratio. 
	Different colors show different cases of training data for mass ratio $q=2$.
	We smooth the data from numerical artifacts using the Savitzky-Golay filter (see text).}
\label{fig:eX}
\end{figure}

As a check, we compute the orbital eccentricity using the 
eccentricity estimator ($e_{\textrm{X}}$) 
and find that the results agree with a maximum relative error of roughly 10\% against 
$e_{\mathrm{comm}}$ quoted in the SXS catalog and given in 
Table.~\ref{tab:sxsdata}.
In Fig.~\ref{fig:eX}, we present the eccentricity estimator $e_{\textrm{X}} 
(X_{\textrm{c}})$ as a function of its circular amplitude and frequency, 
$\mathcal{A}_{\textrm{c}}$ and $\omega_{\textrm{c}}$, respectively.

\subsection{Fitting $e_{\textrm{X}}$}
\label{sec:fitting}

Our main goal is to model an eccentric waveform by modulating the 
amplitude and phase of a circular model.  
To construct the model, we interpolate the additional oscillation of an 
eccentric waveform depending on its eccentricity and mass ratio,
where the relationship between the circular and the eccentric model is expressed in Eq.~(\ref{eq:estimator}).
Accordingly, we look for a fitting function to model $e_{\textrm{X}} (X_{\textrm{c}})$ that relies on the desired parameters ($q$, $e$)
and reverse Eq.~(\ref{eq:estimator}) to obtain the eccentric amplitude and frequency.
We then integrate the frequency to obtain the eccentric phase and construct the eccentric $h_{\mathrm{22}}$.

We note that alternatives to fitting 
the amplitude and frequency modulations have been studied in Ref.~\cite{PhysRevD.103.064022}. In particular, they investigated using
the phase residual instead of the frequency, or fitting the eccentric amplitude and phase (or frequency) directly instead of recasting 
the problem in terms of differences to noneccentric signals. 
Here we find that the most suitable strategy for our approach is to fit the residual amplitude and frequency oscillation defined as the eccentricity estimator ($e_{\textrm{X}}$) that comes from \{$\mathcal{A}_\textrm{22}$, $\omega_\textrm{22}$\} and integrate $\omega_\textrm{22}$ to obtain the phase ($\Psi_\textrm{22}$).

In a suitable parametrization, the eccentricity estimator $e_{\textrm{X}}$ is 
a decaying sinusoidal function (see Fig.~\ref{fig:eX})
with its amplitude defined by the orbital eccentricity $e$
\cite{PhysRevD.82.124016}. 
To model $e_{\textrm{X}}$ for various eccentricities and mass ratios, we fit 
$e_{\textrm{X}}$ with a set of free parameters modifying a damped sinusoidal 
function.
These parameters are two amplitude quantities ($A$ and $B$), a frequency ($f$), 
and phase ($\varphi$) with the following relation:
\be
\label{eq:model}
	e_{\textrm{X}} (X_{\textrm{c}}) = A e^{B \, X_c^\kappa } \sin(f \, 
X_c^\kappa+\varphi).
\ee
$A, B, f$, and $\varphi$ are standard damped sinusoidal parameters obtained from 
the optimized curve fitting.

We use a $X^\kappa_c$ instead of $X_{\textrm{c}}$ to describe the evolution of 
the residual oscillations of the amplitude and frequency mainly for the 
following reasons:
$X_{\textrm{c}}$ is a rapidly evolving function. 
Therefore, it is more difficult to model $e_X$ with a standard sinusoidal 
function with a constant frequency.
Although it is in principle possible to use $X_{\textrm{c}}$ directly in the 
model, we would have to slice the data into multiple small time windows that 
overlap. 
Thus, the results will be less smooth; one would have to blend all those individual functions defined on small intervals into one big function.
Besides, we cannot guarantee our result beyond our calibration range, especially 
for the early inspiral. Using a power law allows us to fit the entire region 
with one set of free parameters.
However, we note that the power law of $X_{\textrm{c}}$ induces a twist 
resulting in infinitely large eccentricities for the very early inspiral stage.
That is a problem with assuming exponential decay, and the fact that the power law we use has a negative exponent. 

We fit our model $e_{\textrm{X}}$ from the starting frequency 
$f_{\textrm{low}}=25\,\mathrm{Hz}$ for a circular \ac{BBH} with a total mass 
$M=50\,M_\odot$.
The power law for $\omega_c$ is $\kappa = -59/24$, and for $\mathcal{A}_c$ it is 
$\kappa = -83/24$.
We emphasize that these values are customized i.e., we expect that one might 
need different values to calibrate with higher eccentricity, a higher mass ratio, 
or a different starting frequency.

\begin{figure*}
	\begin{minipage}{0.49\linewidth}
		\includegraphics[width=\linewidth]{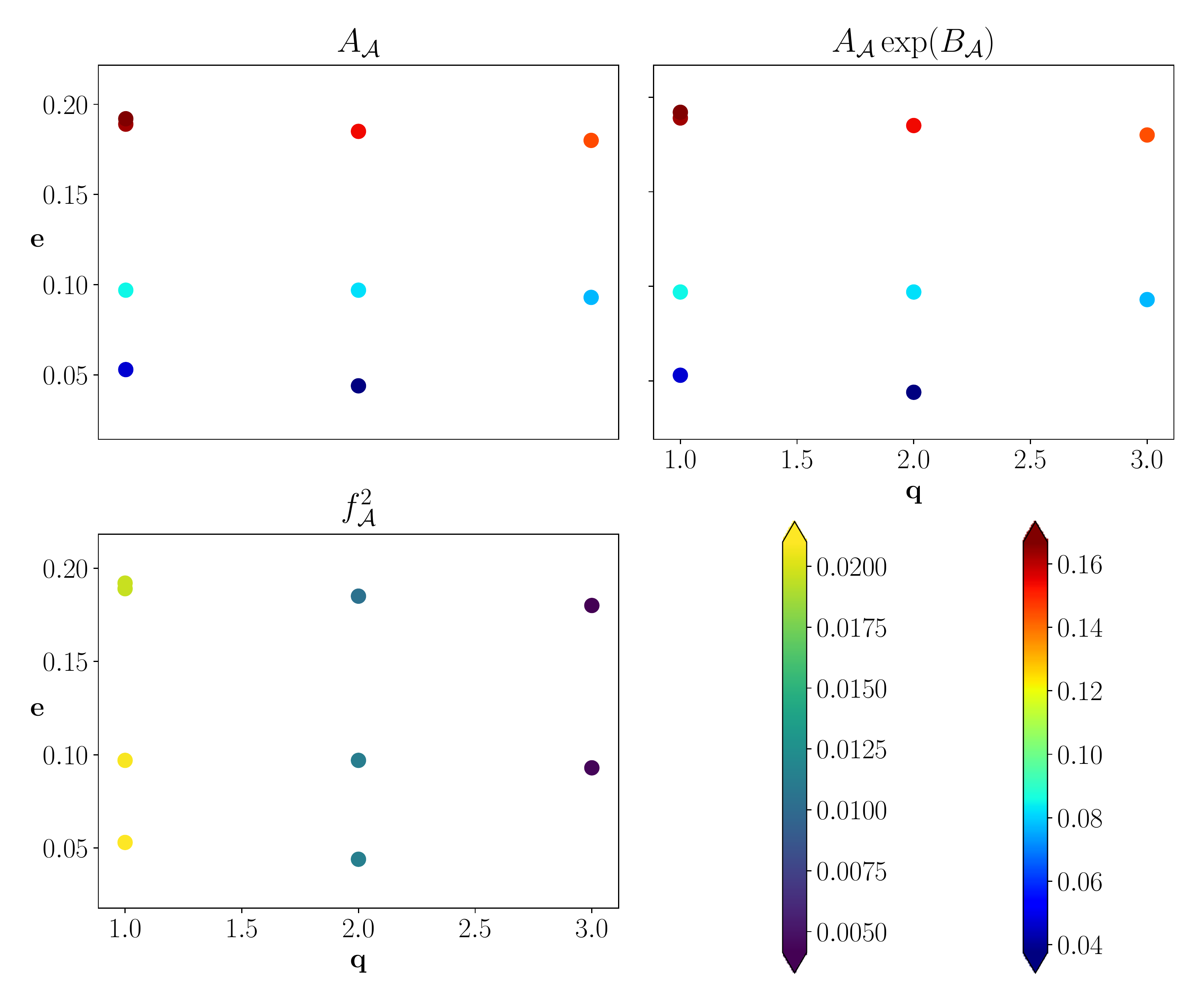}

	\end{minipage} \hfill
	\begin{minipage}{0.49\linewidth}
		\includegraphics[width=\linewidth]{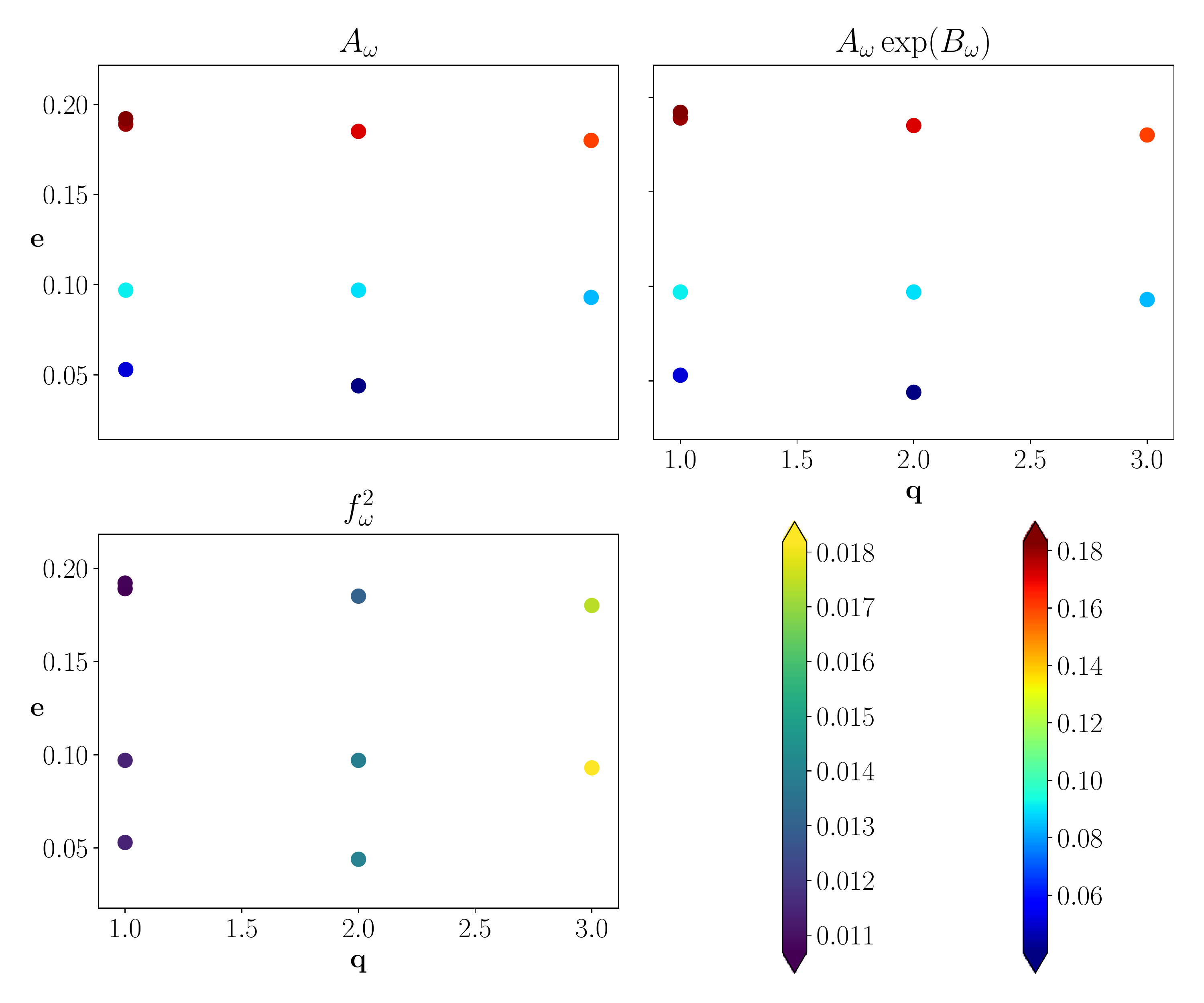}

	\end{minipage}
	\caption{Key quantities of $\mathcal{A}_{\textrm{22}}$ (left) and 
$\omega_{\textrm{22}}$ (right) of a damped sinusoidal function obtained from the 
curve fitting [see Eq.~\ref{eq:model}]. The amplitude parameters ($A$ and $B$) 
depend strongly on the eccentricity ($e$), whereas the square of the frequency 
($f^2$) is correlated to the mass ratio ($q$). We leave $\varphi$ as a free nuisance 
parameter that we maximize over when comparing to the test data. The left color bar corresponds to the bottom panel, and the right color bar to the top panel.}
\label{fig:key_quants}
\end{figure*}

By optimizing the curve fit between $e_{\mathrm{X}}$ and Eq.~\ref{eq:model}, we obtain the four quantities ($A, B, f, \varphi$) for all training data.
The relation between the mass ratio ($q$), eccentricity ($e$), and the three 
parameters $A$, $B$, $f$ is shown in Fig.~\ref{fig:key_quants}.
The amplitude components $A$ and $B$ are strongly correlated to eccentricity, whereas the mass ratio determines the frequency squared.
Hence, we perform one-dimensional linear interpolation across eccentricity to obtain the values of $A$ and $B$. 
Similar to that, we linearly interpolate $f^2$ across mass ratios.
We choose $f^2$ instead of $f$ because the data is smoother for interpolation. 
The square root of $f^2$ gives either positive or negative values. However, this ambiguity can be absorbed by the phase parameter $\varphi$. 

The phase parameter $\varphi$ is an additional degree of freedom that we cannot 
explore sufficiently with the available \ac{NR} data. 
For small sets of \ac{NR} simulations with nearly constant values of $q$ and $e$, but varying $\ell$, we find that the best-fit $\varphi$ 
mirrors changes in $\ell$.
Thus, we expect that it may correlate 
strongly with the mean anomaly. 
Because the orientation of the ellipse is 
astrophysically less interesting than the value of the eccentricity, we do not 
attempt to model the effect of varying the mean anomaly other than introducing 
the phenomenological nuisance parameter $\varphi$. 
We interpolate the other parameters when generating a new waveform 
model with different mass ratios and referenced eccentricities. 

We apply a one-dimensional interpolation for each key quantity shown in Fig.~\ref{fig:key_quants}.
$A$ and $B$ are interpolated over different eccentricities $e$, $f^2$ is 
interpolated over the mass ratio $q$, and the phase of the oscillation 
$\varphi$ can be chosen arbitrarily.

Once we obtain the eccentricity estimators $e_{\textrm{X}}$ using the interpolated quantities, 
we substitute the results to reconstruct $\mathcal{A}_{\textrm{22}}$ and $\omega_{\textrm{22}}$ using Eq.~\ref{eq:estimator}.
To construct $\Psi_{\textrm{22}}$, we integrate $\omega_{\textrm{22}}$ numerically using the trapezoidal rule.
We truncate the waveform at $t=-50M$ and join it with the nonspinning circular model. 
We then smooth the transition with the Savitzky-Golay filter at $-46M<t<-25M$.

We then build h$_{2 \pm 2}$ as the combination of the amplitude and phase as follows:
 \be
\label{eq:strain}
	h_{\ell m}=\mathcal{A}_{\ell m} \; e^{-i \Psi_{\ell m}}.
\ee
To reconstruct the gravitational-wave strain $h=h_{+}-h_\times$, we compute the spin-weighted spherical harmonics $Y_{\ell m}(\iota, \phi)$ and
employ Eq.~\ref{eq:NRform}.


\section{Results}
\label{sec:result}

We built a new nonspinning eccentric model by modulating the residual amplitude and phase oscillations of the circular analytical models, {\texttt{IMRPhenomD}} \cite{PhysRevD.93.044007} and {\texttt{SEOBNRv4}} \cite{PhysRevD.95.044028}.
{\texttt{IMRPhenomD}} is an aligned-spin \ac{IMR} model that was originally 
built in frequency domain and calibrated to numerical simulations for mass 
ratios $q \leq 18$.
{\texttt{SEOBNRv4}} is an aligned-spin time-domain \ac{IMR} model 
\cite{PhysRevD.95.044028,PhysRevD.89.061502} that has been calibrated to 140 
\ac{NR} waveforms produced with the SpEC code up to mass ratio 8 and 
extreme-mass-ratio signals.

As described in Sec.~\ref{sec:method}, we interpolate the residual amplitude 
and phase oscillations of the training dataset for the given mass 
ratio and eccentricity. 
To construct a new, eccentric waveform for the intermediate to near-merger 
regime, we then use 
one of the nonspinning circular models with the desired mass ratio, compute 
the eccentricity estimators ($e_X$) from the analytical description given in 
Eq.~(\ref{eq:model}), and reconstruct the desired eccentric waveform model for 
each test data. 
We develop a map from circular nonspinning waveforms to eccentric waveforms that can be applied to any analytical model with a relatively simple and fast function using only 20 \ac{NR} simulations.

We evaluate the results by computing the overlap between the new model and the 
\ac{NR} test data. The overlap is maximized over a time and phase shift, as 
well as the free phase offsets of the residual oscillations. 
Mathematically, we define the overlap $\mathcal O$ based on an inner product between two waveforms:
\begin{align}
\langle h_1, h_2 \rangle  & = 4 \operatorname{Re} \int_{f_1}^{f_2} \frac{\tilde{h}_1(f) \, \tilde{h}_2^* 
(f)}{\mathrm{S_n}(f)} \mathrm{d}f, \\
\mathcal{O} &= \max_{\{t_0, \Psi_0, \varphi_\mathcal A, \varphi_\omega \}} 
 \frac{\langle h_1, h_2 \rangle  }{\|h_1\| \|h_2\|},
\end{align}
where $\mathrm{S_n}$ is the sensitivity curve of the corresponding \ac{GW} 
interferometer, $\tilde h(f)$ is the Fourier transform of $h(t)$, $^\ast$ denotes complex conjugation and $\| h \| = \sqrt{\langle h, h \rangle}$.
The mismatch or unfaithfulness is defined by
\be
\mathcal{M}= 1-\mathcal{O}.
\ee

We investigate three sensitivity curves for the future \ac{GW} 
detectors, \ac{aLIGO} A+, the \ac{ET}, and \ac{CE}.
LIGO A+ is the future \ac{GW} interferometer with 1.7 times better 
sensitivity than the current detector, expected to start observing in 
mid-2024 at the earliest \cite{NSFPressConf}.
The \ac{ET} is a 10 km \ac{GW} observatory planned to be built on the border between Belgium, Germany, and Netherlands which could be operating in the mid-2030s \cite{Maggiore_2020}. 
The \ac{ET} is expected to have higher sensitivity towards the low-frequency range.
\ac{CE} is a 40 km third-generation \ac{GW} detector which has higher sensitivity towards low redshift ($z>10$) that is planned to start observing in the 2030s \cite{CEdocs}.
Since our model focuses on the late inspiral case, and because 
the unfaithfulness is insensitive to a change in overall signal-to-noise ratio, 
the values obtained for the future third-generation detectors show similar 
behavior \cite{thirdgen}.
Hence, we only show the overlap results for the LIGO A+ design sensitivity.
A possible caveat is that our model might not fill the LIGO A+ band down to 10 Hz. 
Thus, there is a chunk of inspiral power missing in the signal. 

\begin{figure*}
\includegraphics[width=\linewidth]{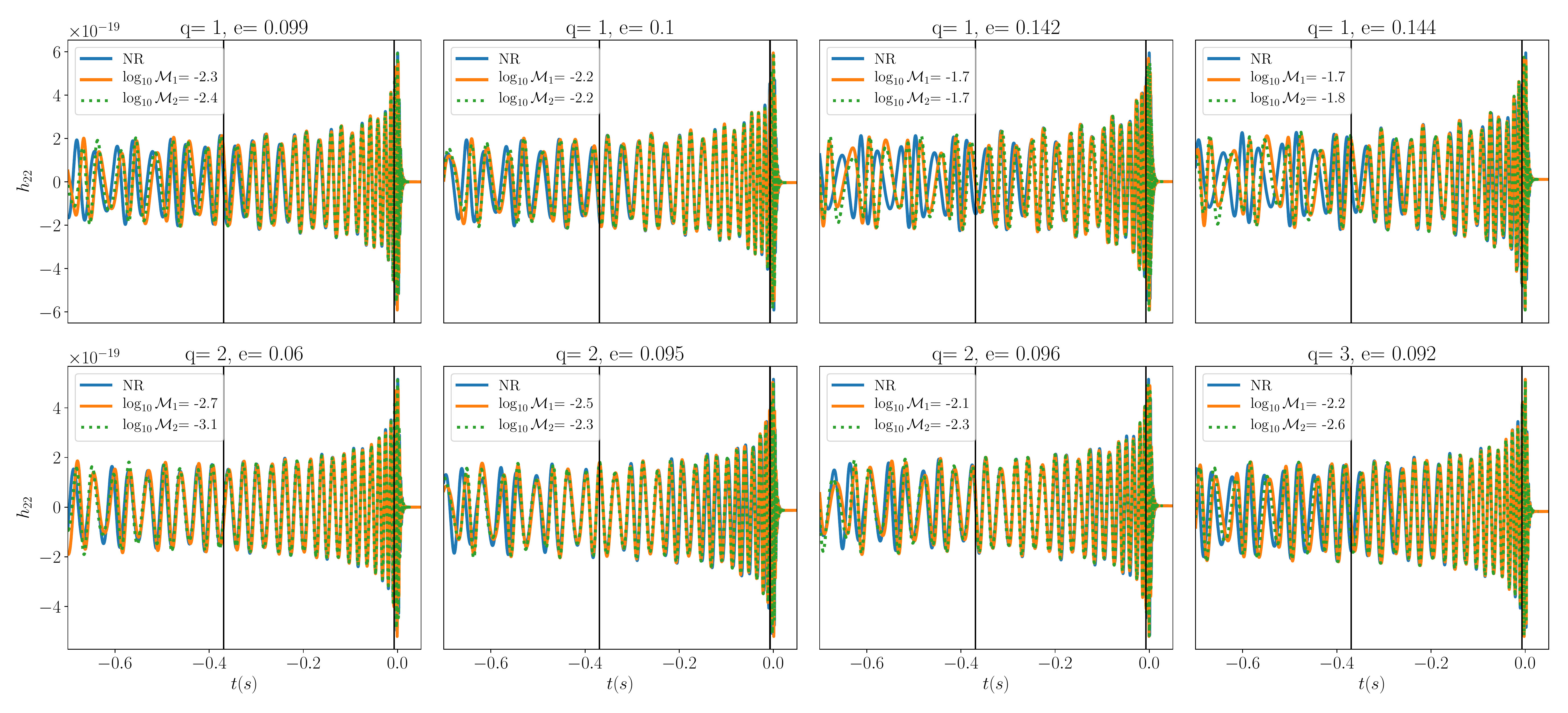}
\caption{{\texttt{IMRPhenomD}} (orange) and {\texttt{SEOBNRv4}} 
(green) circular waveforms twisted into eccentric models. $\log_{10} 
\mathcal{M}_1$ is the log mismatch of {\texttt{IMRPhenomD}} against the \acp{NR} 
waveform (shown in blue), and $\log_{10} \mathcal{M}_2$ gives the log mismatch 
of {\texttt{SEOBNRv4}} 
against \ac{NR} with the same mass ratio and eccentricity, respectively. 
The total 
mass of the system is $M\,=\,50 M_\odot$, and the mass ratio ($q$) and eccentricity 
($e$) are shown in the title of each plot. We employ the A+ design sensitivity 
curve starting at $f\,=\,35\,\textrm{Hz}$ (see text) to compute the match. The 
black 
vertical lines mark the range in which we perform the interpolation and 
compute the match.}
\label{fig:modelresults}
\end{figure*}

Figure~\ref{fig:modelresults} visually compares the strain $h_\mathrm{2 \pm 2}$ 
of each \ac{NR} test dataset with the new eccentric nonspinning signal built from 
analytical models, {\texttt{IMRPhenomD}} and {\texttt{SEOBNRv4}} 
for a $50\,M_\odot$ \ac{BBH} with inclination angle $\iota$=0 (face-on) and phase of coalescence, $\phi_c$=0.
Using our method, we find that the minimum overlap between the new model and \ac{NR} is $\approx 0.98$ ($\log_\textrm{10}\mathcal{M}\,=\,-1.8$) over all of our test datasets.
The minimum overlap occurs at the highest eccentricity in the test dataset.

\begin{figure*}
	\begin{minipage}{0.49\linewidth}
		\includegraphics[width=\linewidth]{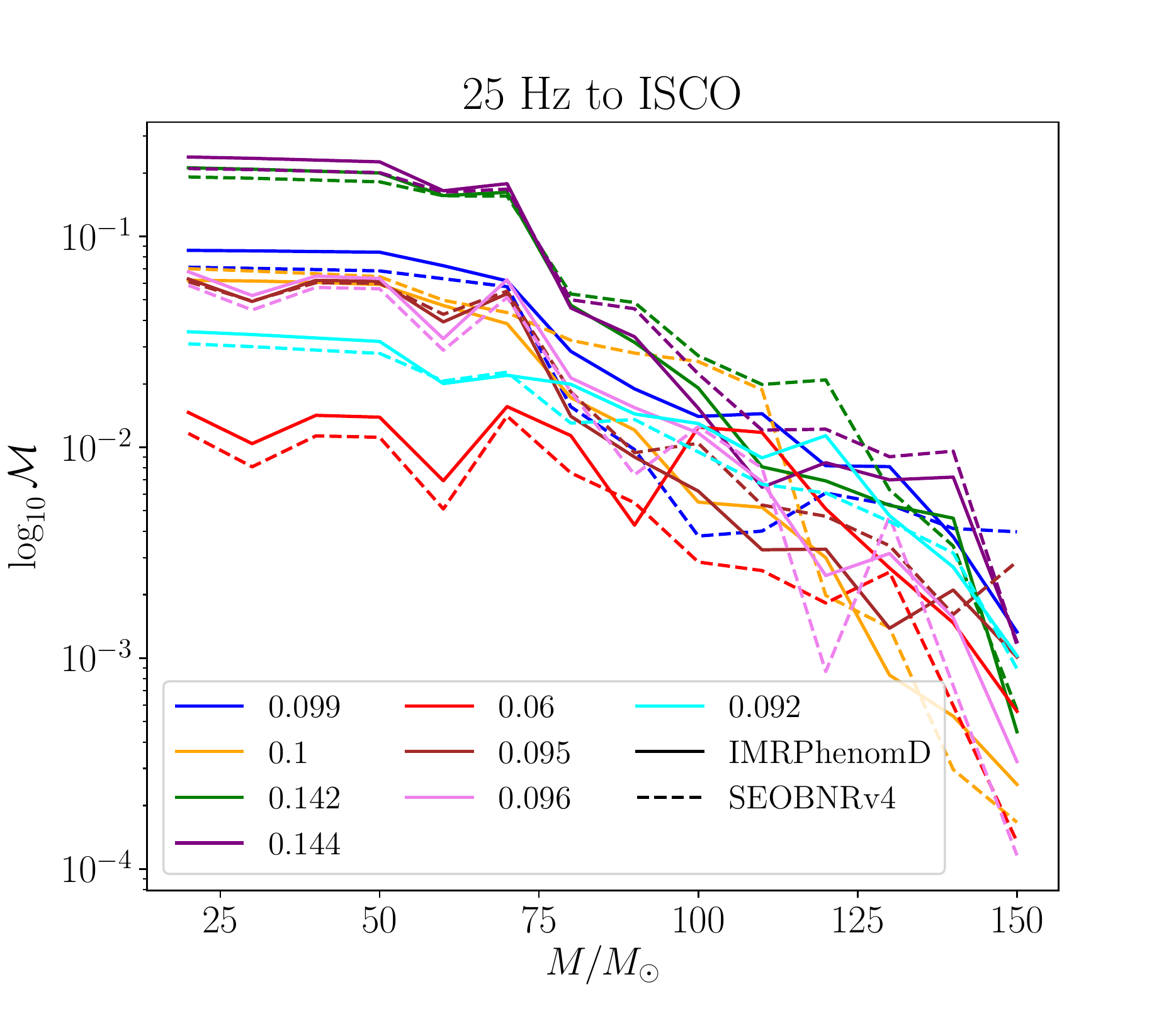}

	\end{minipage} \hfill
	\begin{minipage}{0.49\linewidth}
		\includegraphics[width=\linewidth]{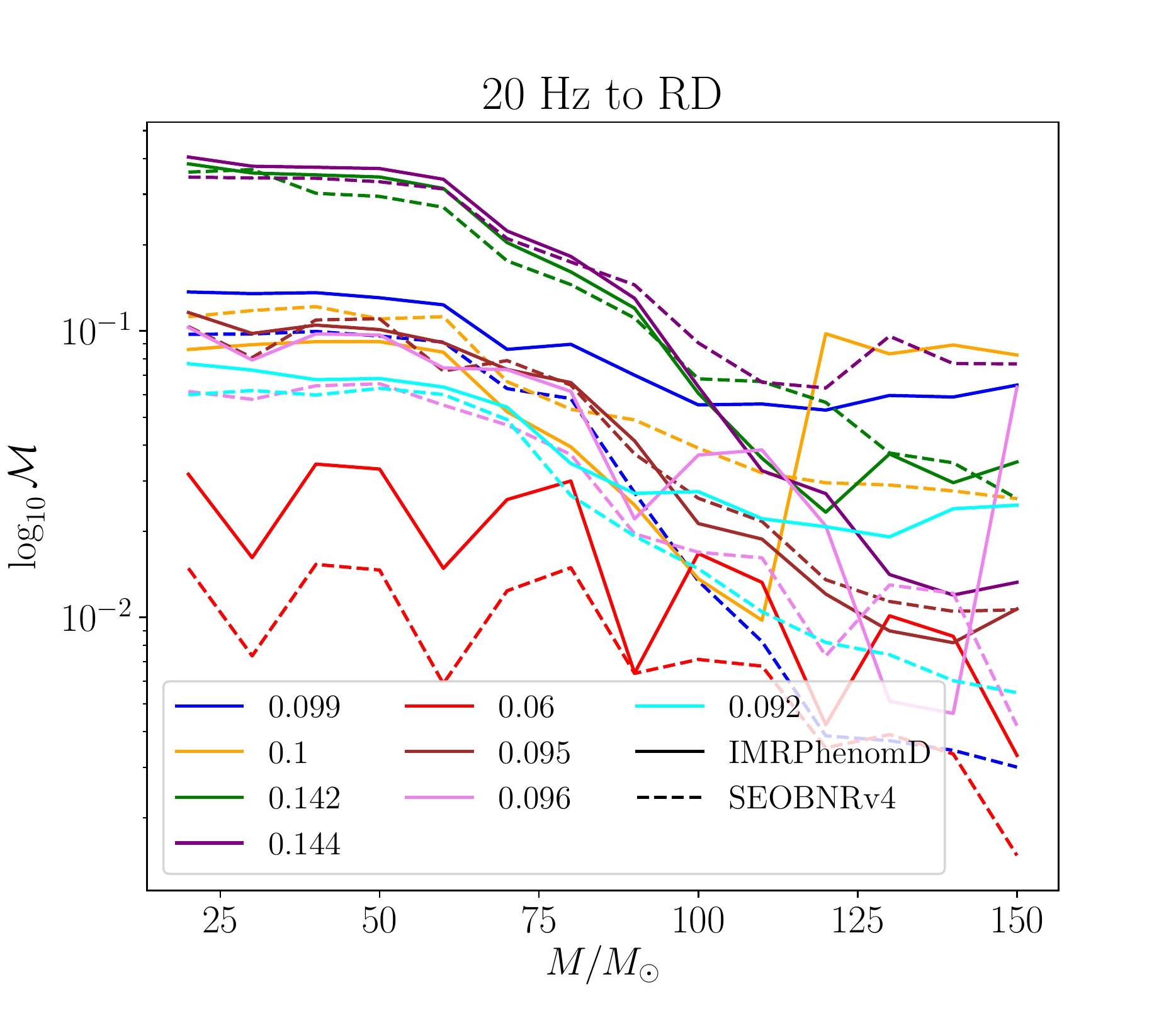}

	\end{minipage}
	\caption{Mismatch results of eccentric variants of {\texttt{IMRPhenomD}} 
and {\texttt{SEOBNRv4}} 
against the \ac{NR} test data for different total masses assuming aLIGO A+ 
design 
sensitivity. 
	Left: $25\,Hz$ to ISCO frequency (within the calibration range). Right: 
from 20 Hz to ringdown frequency (beyond the calibration range), where we define 
the ringdown frequency as $f_\textrm{RD}$=$0.1/ M$.}
\label{fig:flowtotalmass}
\end{figure*}

Although we calibrated the new model for limited ranges in mass ratio, 
eccentricity, and time, we let the production of the new model go beyond our 
calibration range.
In Fig.~\ref{fig:flowtotalmass}, we show the unfaithfulness of the new model against the \ac{NR} test data for various total masses with the aLIGO A+ design sensitivity curve.
The left panel shows the unfaithfulness within the calibrated frequency range, between 25 Hz and the ISCO frequency scaled over the total mass.
Similarly, the right panel presents the unfaithfulness beyond the calibrated frequency range, between 20 Hz and the ringdown frequency.
We use the definitions of the ISCO and ringdown frequencies as follows:
\be
\label{eq:fISCO}
f_\textrm{ISCO}=1/(6^{3/2} \pi M),
\ee
and
\be
\label{eq:fRD}
f_\textrm{RD}=0.1/M.
\ee

Figure~\ref{fig:flowtotalmass} shows that the mismatches decrease toward higher-total-mass systems.
As the total mass increases, the overlap computation covers a smaller waveform regime towards merger in the frequency space.
Since the eccentricity decreases over time, the near-merger regime has lower eccentricities. 
Thus, the overlap between the model and the corresponding \ac{NR} simulation is better for the higher-mass systems compared to the lower-mass ones.
For comparison, we find that mismatches between circular analytical models and the eccentric \ac{NR} test data are at least 1 order of magnitude worse than the results we find for our eccentric model.

The unfaithfulness between eccentric waveforms is better for \{25, ISCO\} than for \{20, Ringdown\}.
We investigate the contribution weight between the early inspiral and the ringdown in the unfaithfulness results by comparing with the \{25, Ringdown\} and \{20, ISCO\} ranges.
We argue that the mismatches for the low masses are dominated by the inspiral, whereas for high masses, the mismatches are dominated by the merger or ringdown.
In the mismatch computation, we add padding in the ringdown area, but the early inspiral should come purely from the fitting data.

\begin{figure*}
\begin{minipage}{0.96\linewidth}
	\includegraphics[width=\linewidth]{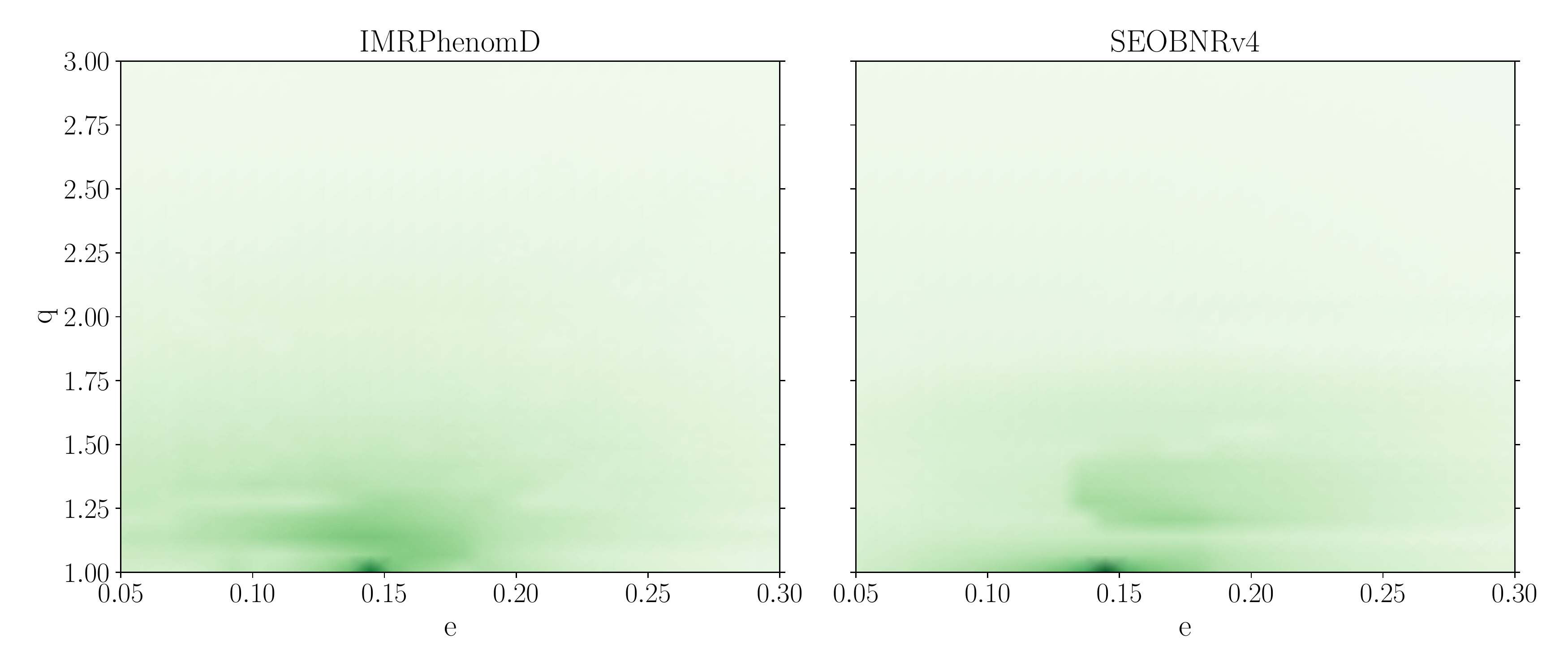}
\end{minipage} \hfill
\begin{minipage}{0.03\linewidth}
	\includegraphics[width=\linewidth]{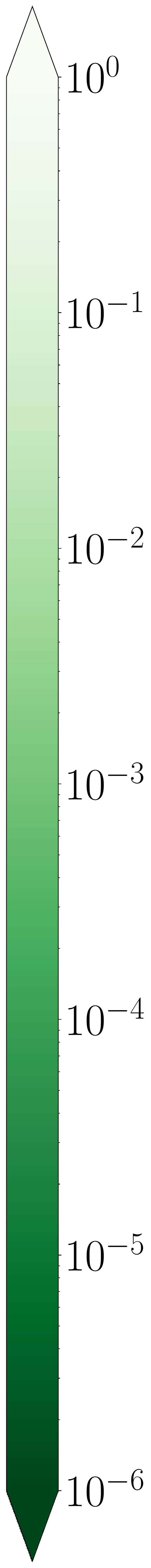}
\end{minipage}
\caption{Comparison with the highest eccentricity in the test dataset, $e=0.144$, $q=2$.
We generate an eccentric waveform model derived from a nonspinning circular 
model, {\texttt{IMRPhenomD}} or {\texttt{SEOBNRv4}}, and compare the signal with 
models for different mass ratios and eccentricities.
Waveforms with higher parameter distance have lower overlap. 
The color bar shows the $\log_{\rm{10}}$ mismatch.}
\label{fig:injection}
\end{figure*}

Furthermore, we test how well one can extract the parameters of an eccentric 
signal $h(q, e)$ by comparing with various waveforms with different 
eccentricities $e$ and mass ratios $q$.
We generate a \pyrex waveform ($q=1$, $e=0.144$) and compare it with various 
other signal parameters ($q, e$) using the same analytical 
waveform model. The results are shown in 
Fig.~\ref{fig:injection}.
We emphasize that in this study, we did not run a standard \ac{PE} pipeline 
that stochastically explores a much greater parameter space. In particular, we 
do not consider varying the total mass or spin. Hence, our results are only a 
first indication of potential parameter ambiguities.
Our results in Fig.~\ref{fig:injection} show that the mismatch between the 
generated waveform and other waveforms having similar mass ratios but different 
eccentricities is relatively low, suggesting that an accurate measurement of the 
eccentricity is challenging for high-mass \ac{BBH} systems where only the late 
inspiral and merger are accessible through the \ac{GW} detection.

\section{Conclusion and future perspectives}
\label{sec:conclusion}
The detection of \acp{GW} from an eccentric \ac{BBH} merger would be a crucial 
step towards understanding the physical evolution of compact binary 
coalescences and the nature of \acp{BBH} in globular clusters. 
Due to limitations in waveform modeling, the current search and parameter 
estimation pipelines in the LIGO/Virgo data analysis rely on analytical 
waveform models for circular binaries.
One of the limitations to developing eccentric \ac{BBH} models is the small 
number of eccentric \ac{NR} simulations.
\ac{NR} simulations that are publicly available have low eccentricities ($e \leq 0.2$) at $M\omega^{2/3}$\,=\,0.075.
We use 20 \ac{NR} simulations from the open SXS catalog and split them into 12 
training datasets and 8 test datasets to develop our method.

We presented a novel method to convert any circular nonspinning waveform model 
into a low-eccentricity nonspinning waveform.
To develop our method, we analyzed the residual modulations in the amplitude 
and frequency of eccentric waveforms compared to the circular signals with the 
same mass ratio in the 12 \ac{NR} simulations of the training dataset.
We modeled the decrease of eccentricity over time, known as the eccentricity 
estimators, $e_X$, using a damped sinusoidal fit,
where the fitting function is built upon four key parameters.
We then performed a one-dimensional interpolation for each key parameter ($A$, 
$B$, and $f$) to build the eccentric waveform with the desired mass ratio and 
eccentricity.
One of our model parameters, $\varphi$, shows no clear correlation with the 
physical parameters we explore. However, the small number of \ac{NR} 
simulations used here did not allow us to model the effect of varying the mean 
anomaly in detail, and we expect $\varphi$ to represent this degree of freedom.
When quantifying the agreement between our model and the test data, we maximize 
over this nuisance parameter.

We then build a new model using the fitting values of $e_X$ and the amplitude 
and frequency of the circular model which here we take from 
{\texttt{IMRPhenomD}} and {\texttt{SEOBNRv4}}.
Our new model has an overlap $0.98 \lesssim \mathcal{O} \lesssim 0.999$ over all \ac{NR} simulations in our test dataset with the LIGO A+ design sensitivity curve. 
We hint that we need more training and test datasets for further development of this model beyond the current parameter boundaries.
The computation of our method can be performed easily and quickly in the Python package \pyrex \cite{pyrexzen}.

Although we calibrate our model to a 50 $M_\odot$ \ac{BBH} ($q \leq 3$, $e \leq 
0.2$) starting at frequency $f_\textrm{low}=25$ Hz, we let the computation 
go slightly beyond the calibrated range.
The calibrated time range of the waveform is from the late inspiral up to the near-merger phase, but we can extend the model through merger and ringdown by 
using the circular data. For the early inspiral, an analytical \ac{PN} model 
could be used to complete the description of the entire coalescence. 
This way, our approach can be adapted to develop a complete \ac{IMR} 
eccentric model.
This would be especially important for future generations of \ac{GW} interferometers 
as they have higher sensitivity especially 
in the low-frequency range. 
Careful studies of eccentric search and parameter estimation are needed to detect eccentric compact binary coalescences and their origin.

\begin{acknowledgments}
The authors would like to thank David Yeeles, Maria Haney, and Sebastian Khan for useful discussions, and the anonymous referee for insightful comments on the manuscript.
Computations were carried out on the Holodeck cluster of the Max Planck Independent Research Group ``Binary Merger Observations and Numerical Relativity'' and 
the LIGO Laboratory computing cluster at California Institute of Technology.
This work was supported by the Max Planck Society's Research Group Grant. 
\end{acknowledgments}
\appendix

\bibliographystyle{vancouver}
\bibliography{manuscript}

\end{document}